\documentclass[Galaxies,article,accept,oneauthor,pdftex,10pt,a4paper]{mdpi} 

\firstpage{1} 
\articlenumber{12}
\doinum{10.3390/galaxies8010012}
\pubvolume{8}
\pubyear{2020}
\copyrightyear{2020}
\history{Received: 18 December 2019; Accepted: 27 January 2020; Published: 8 February 2020}

 \theoremstyle{mdpi}
 \newcounter{thm}
 \newcounter{ex}
 \newcounter{re}
 \theoremstyle{mdpidefinition}

\newcommand{\dd}{\textnormal{\,d}}
\newcommand{\erf}{\textnormal{\,erf}}
\newcommand{\hh}{\textnormal{H\textsubscript{I}}}

\Title{Entropy and Mass Distribution in Disc Galaxies}

\Author{John Herbert Marr $^{1}$}
\AuthorNames{John Herbert Marr}

\address{%
$^{1}$ \quad Unit of Computational Science, Building 250, Babraham Research Campus, Cambridge, CB22 3AT, UK; 
john.marr@2from.com}

\abstract{
The relaxed motion of stars and gas in galactic discs is well approximated by a rotational velocity that is a function of radial position only, implying  that individual components have lost any information about their prior states. 
Thermodynamically, such an equilibrium state is a microcanonical ensemble with maximum entropy, characterised by a lognormal probability distribution.
Assuming this for the surface density distribution yields rotation curves that closely match observational data across a wide range of disc masses and galaxy types, and provides a useful tool for modelling the theoretical density distribution in the disc.
A universal disc spin parameter emerges from the model, giving a tight virial mass estimator with strong correlation between angular momentum and disc mass, suggesting a mechanism by which the proto-disc developed by dumping excess mass to the core, or excess angular momentum to a satellite galaxy.
The baryonic-to-dynamic mass ratio for the model approaches unity for high mass galaxies, but is generally $<1$ for low mass discs, and this discrepancy appears to follow a similar relationship to that shown in recent work on the radial acceleration relation (RAR). 
Although this may support Modified Newtonian Dynamics (MOND) in preference to a dark matter (DM) halo, it does not exclude undetected baryonic mass or a gravitational DM component in the disc. 
}

\keyword{galaxies: kinematics and dynamics; galaxies: spiral; lognormal density distribution; galaxies: entropy}

\begin{document}

\section{Introduction}
\label{Intro}
In  classical  thermodynamics,  entropy  is  defined  as  a  state  function  of  the  system.   
This  is  a property  that -- like volume, density and internal energy -- is  dependent  only  on  the  current  state  of  the system  and  independent  of  how  that state was achieved, being in principle measurable to any precision. 
As with any thermodynamic system, cosmological systems such as a galaxies typically  have  a  large  number  of  particles including gas  molecules, dust and stars, in any of a huge number of possible arrangements within the system defining the amount of information needed to specify the state of the system as a measure of its entropy. 

\citet{1969ApJ...155..393P} postulated that galactic spins originated from induced tidal torques from neighbouring structures, whereby initially spherical galaxies developed their pronounced disc shape through acquired angular momentum. 
The initial spin impulse would have added kinetic energy to the system through induced movement, and potential energy through drawing out filaments of stars from merging galaxies. 
More recently, \citet*{2017MNRAS.467.5022H} proposed that, with the addition of torque, radial migration effectively mixes the angular momentum components of a proto-galaxy to produce the observed circular orbits while conserving total mass and angular momentum, such that the disc's distribution of specific angular momenta $j$ should be near a maximum entropy state.

The stability and ubiquitous nature of galactic discs imply that any relative motion of a displaced star will rapidly be partitioned on the disc rotational time-scale. 
\citet{2002MNRAS.336..785S} have demonstrated that the spiral waves in galaxy discs provide an effective method of radial mixing by churning the baryonic components in a manner that preserves the overall angular momentum with little increase in random motion, and on observational grounds there is generally little tendency for macroscopic turbulence to occur within the stable disc.
This does not preclude local gravitational interactions producing other parameters of velocity distribution, such as the age-velocity dispersion relation described by \citet*{2004MNRAS.350..627D} in the solar neighbourhood, or the formation of local areas of over- or under-density such as spiral arms, bars or voids, and the influence of these on chaotic motion in the solar region has been explored by \citet{2009ASSP....8..151C}.
There are therefore strong analogies between this additional energy, and the addition of a volume of hot gas into a chamber of cold gas with the mixture allowed to diffuse to equilibrium; and the relaxed state of the disc has parallels with adiabatically isolated thermodynamic systems of interacting particles that attain statistical equilibrium with an increase of entropy.

Galactic discs may not be fully isolated as they can eject high-energy stars and may acquire some loose stars and gas from other systems, but once formed they are thought to have had minimal active development over the past 6 Gyr \citep{2000ARep...44..711T, 2010AJ....140..663G, 2016NatAs...1E...3P}, and even the massive Brightest Cluster Galaxies (BCGs) have shown only slow changes in overall brightness over the last 3.5 Gyr of their 10 Gyr of observable history \citep{2016MNRAS.460.2862B}, suggesting that external accretion is not required to sustain star formation.
Over the recent past, discs may therefore be considered as isolated assemblies of discrete particles undergoing only conservative gravitational interactions, with conservation of independent parameters such as total angular momentum, internal energy, overall number of baryons, and total disc mass.
In addition to the conserved macroscopic variables, the observed stable circular orbits imply that all information about the initial conditions has been lost.
This implies that reversing the orbits will produce a mirror version of the disc, but will not recreate the original proto-galaxy.
On a macro scale, such a system approximates an adiabatic thermodynamic system in thermal equilibrium with a time-invariant, stable mass distribution.
This defines a microcanonical ensemble whose entropy is maximized through chaotic mixing of its components as the system approaches equilibrium. 

A dynamical approach to describe the relaxation process is difficult as no exact description of the initial state is known, but the similarity and stability of disc galaxies allows them to be considered as idealized relaxed systems in an equilibrium state. 
The typical disc has $\sim 10^{8}-10^{12}$ stars, and with large $N$ it is necessary to consider the average, statistical properties of the system rather than individual orbits.
In contrast to systems with short range repulsive interactions like neutral gases, the attractive long range gravitational force precludes the standard methods of statistical mechanics from being used directly, but the microcanonical distribution can be used to study the statistical properties of any closed system with a fixed total energy $E$, and gravitating systems can also be described by this distribution \citep{1990PhR...188..285P}.
Such a relaxed state has parallels with adiabatically isolated thermodynamic systems of interacting particles that attain statistical equilibrium with increase of entropy and an associated mass distribution that is essentially lognormal (LN).
Hence interpreting the disc as a thermodynamic system with high entropy also provides a model universal distribution function for rotationally supported discs with a minimum of free parameters  \citep{2015MNRAS.448.3229M, 2015MNRAS.453.2214M}.
By looking at disc galaxies as isolated systems with maximal entropy, the theoretical dynamic masses can be computed for a wide range of galaxies and compared with their observational masses. 
We discuss how such systems may have developed their observed mass-density distributions in the disc, and compare these with observations to consider potential mechanisms to describe them.

\section{Entropy changes within an evolving galaxy}
\label{Entropy}
Several papers have considered the role of entropy in galaxy formation and structure, and entropy optimization provides a powerful method for data analysis \citep{2008Binney}.
A maximum entropy approach has been utilized to describe the local structures of the velocity distribution for the phase density function of several samples from the HIPPARCOS and Geneva-Copenhagen survey catalogues \citep{2005AAp...442..929A}, while \citet{2010AAp...510A.103C} concluded that the entropy method offers an excellent estimation of the truncated velocity distributions of samples containing only thin disc stars. 

Following Peebles' conjecture \citep{1969ApJ...155..393P}, assume a filament of mass $\delta M$ to be drawn out from a galaxy of total mass $M$. In the absence of torque, the filament will provide an added gravitational potential $\delta E$ to the system, such that the total internal energy $E$ of the system increases.
The filament will then fall back into the body of the galaxy such that the potential energy of its individual components will convert to kinetic energy with added radial velocity that will equilibrate over time through mechanisms such as dynamically important strong galactic magnetic fields.
These provide the transport of angular momentum required for the collapse of gas clouds and the formation of new stars, drive mass inflow into the centres of galaxies, and can affect the rotation of gas in the outer regions of galaxies, playing important roles in the evolution of galaxies through their direct impact on star formation and stellar feedback-induced turbulence \citep{Beck:2007, 2018A&A...619L...5N}.
This results in a large number of individual stars acquiring additional radial energy, and the volume of the galaxy will expand.
The transfer of torque to the galaxy will also transfer kinetic energy of rotation to the system, adding to the total internal energy.
The total disc energy (kinetic plus potential) is then the sum of the individual energies:
\begin{equation}
E=\sum_{i=1}^{N}\epsilon(i)\,.
\label{eq:Utot}
\end{equation}

A system of $N$ particles can be described at any moment by a point in a phase space of $6N+1$ dimensions, parametrized by the 6$N$ canonical co-ordinates and momenta, and time ($q_i$,~$p_i$,~$t$). 
These 6$N$ functions are conserved, and at equilibrium the statistical behaviour becomes independent of time, therefore $P(q, p, t)\approx P(q, p)$, with the phase point tracing a one-dimensional curve on a hypersurface of constant $E$ in phase-space \citep{1990PhR...188..285P}. 
Observations appear to confirm that the equilibrium behaviour of a smoothly rotating galactic disc is independent of its initial conditions \citep{2008Binney}, obeying statistical regularity in conformity to the microcanonical distribution. The average for any phase-space function $f(p,q)$ is then given by Eq.~\ref{eq:phasespace}: 

\begin{equation}
\langle f(p,q)\rangle = \left( \frac{1}{N! g(E) } \right) \int{f(p,q)}\delta (E-H(p,q)) \dd{p}\dd{q}\,,
\label{eq:phasespace}
\end{equation}
where $g(E)$ is the density of states, $N!g(E)$ is the volume of phase-space and $H(p,q)$ is the Hamiltonian \citep{1990PhR...188..285P}.

As with all classical systems, because the properties are continuous, the number of microstates is uncountably infinite. 
The microstates must therefore be grouped by a coarse graining technique by defining their positions and momenta within limited ranges of volume and momenta, $\delta{V}$ and $\delta{p}$, to obtain a countable set to define $g(E)$ \citep{1965fstp.book.....R}. 
For microscopic systems, these limits are set by quantum parameters.
For a macroscopic system such as a galactic disc, it is sufficient to define $\delta{V}$ in terms of a capture volume, and $\delta{p}$ in terms of a differential rotational momenta such that individual pairs of stars remain separated over a timescale that is long when measured against their periods of rotation around the disc.

The gravitational potential of a star of 1 solar mass at a distance of 4~light years from the sun is $U\approx -3.78\times 10^3$~J/kg, in contrast to that for the Milky Way disc at the position of the sun ($\approx 8$~kpc from the centre) of $U\approx -2.68\times 10^{11}$~J/kg , a factor of $7.1\times10^7$, and to have a comparable acceleration to the disc, a star of 1~solar mass would have to enter the Oort cloud ($\sim$5000~A.U.). 
Although a gross simplification, this does justify considering close encounters to be weak interactions.
Standard arguments then show that if $N\gg 1$ and the interactions are weak, the relative probability  associated with the distribution function $\langle f(p,q)\rangle$ is a distribution of particles in N-dimensional space that is uniform on the manifold. 
Although the motion of individual stars in a many-body gravitational field is chaotic, with no possibility of recreating the original state of motion of the galaxy at its formation, observations confirm that the equilibrium behaviour of the rotating galactic disc obeys statistical regularity.
Because the volume, mass, and internal energy are fixed at equilibrium (the microcanonical ensemble), such a system is one in which all states are equally likely and independent of the initial conditions.
Maximisation of the entropy is then equivalent to maximising the phase-volume, with the entropy of the system $S(E)$ given by Eq.~\ref{eq:SE}: 

\begin{equation}
S(E) = K_s\ln{(g(E))}\,.
\label{eq:SE}
\end{equation}
where $K_S$ is a dynamical constant of the system, analogous to $K_B$, the Boltzmann constant in thermodynamics.

For the disc system, a more turbulent state may be assumed to have existed at the formation time of the galaxy, when components had acquired angular momentum but not settled into a regular disc.
Such an early structure would have a unique position and momentum signature for each component, and -- perhaps counter-intuitively -- turbulent motion is a more ordered state than laminar flow, and hence the transition towards laminar flow is accompanied by an increase in entropy \citep{2016Sieniutycz}. 
Once the disc has stabilized, the gross motion of its components is described by the radial variable, which represents a reduction in information about the system as a whole.
An appropriate model for the flow of stars in the galactic disc is suggested by a hydrodynamic analogy to the adiabatic laminar flow of fluid through a thin, flat pipe. Laminar flow is a flow regime characterized by low-momentum convection but high-momentum diffusion, and these conditions may be extended to the disc components moving in approximately circular orbits at constant linear velocity, with a radial differential velocity across the disc. 
The stability and ubiquitous nature of galactic discs suggests that any relative motion of a displaced star will rapidly be dissipated---compared with the  time-scale of disc rotation---to allow it to match the specific momentum of gravitational mass at its new position \citep{2017MNRAS.467.5022H}. 
Detailed analysis, however, remains complex even within a well-defined hydrodynamic system such as laminar flow through a circular pipe \citep{2013JThermo?2013?234264}.

We seek to maximise entropy using the Wallis probability distribution \cite{2003Jaynes}.
Consider a system with $n$ mutually exclusive states, assigned probabilities $(p_1,p_2,\cdots, p_n)$, and let $q$ quanta of probability, each worth $\delta=1/q$, be randomly distributed among the $n$ possibilities. 
Then $p_i=q_i/q$, where $p_i$~is the probability of the $i^{th}$ position $(i=1,2,\cdots, n)$ and $q_i$~is the number of quanta assigned to the $i^{th}$ position, with
\begin{equation}
    \sum_{i=1}^{n} p_i=1\,,
\end{equation}

The probability of $p$ is the multinomial distribution \cite{2003Jaynes}:
\begin{equation}
    P_r(p)=n^{-q}.W\,,
\label{eq:Pr}
\end{equation}
\begin{equation}
    \textnormal{where}~~~W=\frac{q!}{q_1!\,q_2!\,\cdots\,q_n!}= \frac{q!}{(qp_1)!\,(qp_2)!\,\cdots\,(qp_n)!}\,.
    \label{eq:W}
\end{equation}

The most probable outcome is then the maximum of $W$. 
We can equally maximise any monotonic increasing function of W and this is most easily achieved by maximising $q^{-1} \log(W)$, using:
\begin{equation}
    \frac{1}{q}\log W=\frac{1}{q}\log \frac{q!}{(qp_1)!\,(qp_2)!\,\cdots\,(qp_n)!}
    \label{eq:W2}
\end{equation}

We may simplify (\ref{eq:W2}) by the Stirling approximation \cite{2003Jaynes}.
Letting the quanta size $\delta\rightarrow 0$ as the number of quanta $q\rightarrow\infty$, the probability levels go from discrete and grainy to smoothly continuous:
\begin{equation}
     \lim_{q\rightarrow\infty}\frac{1}{q}\log(W) \rightarrow -\sum_{i=1}^n p_i \log(p_i)\,,
     \label{eq:S2}
\end{equation}
and it may be shown that, for large $n$, (\ref{eq:S2}) is the entropy of the system \cite{2003Jaynes}: 
\begin{equation}
      -\sum_{i=1}^n p_i \log(p_i)=S(p_1,p_2,\cdots,p_n)\,.
      \label{eq:S3}
\end{equation}

Therefore maximising (\ref{eq:S2}) is equivalent to maximising the entropy \cite{2003Jaynes}. 
For a normal distribution of $(p_1,p_2,\cdots,p_n)$, (\ref{eq:S2}) will have a lognormal distribution, which is the distribution of the product of independent random variables \citep{2001BioScience..51..341}.
By equating (\ref{eq:S3}) with (\ref{eq:SE}), the lognormal distribution is then the maximum entropy probability distribution for the disc system \citep{2009JEconometrics..150..219}, which is the canonical ensemble, originally derived by Gibbs as the maximum entropy distribution over the classical state space, or phase volume, based on a specified mean value of the energy.

\subsection{Rationale for a Lognormal Density Distribution}
\label{Sec:LNDist}
The ability to build an accurate model is helpful when comparing the theoretical dynamical mass with the observational mass, and a number of methods have been described to generate a disc mass density distribution that mimics any observed rotational curve (RC), generally by using {\em ad hoc} fitting models.
Because the disc is thin compared to its radius, most analytical studies assume it to have negligible thickness and describe it in terms of a pure surface density function, $\Sigma(r)$. 
This is generally taken to be a function of the surface brightness of the disc, and on empirical grounds this led to a model of the disc as an exponential function of the form $\Sigma(r)=\Sigma_0 \exp{(-r/r_0)^{1/n}}$ where $r_0$ is a characteristic scale factor for the galaxy, with $0.5\leq{n}\leq 10$ (the S\'{e}rsic function) and many simulations of disc formation lead to an approximation of this exponential form with $n=1$ for the galactic disc (the {\it Freeman disc}) \citep{2017MNRAS.467.5022H, 2013MNRAS.428..129S}.

The unification of thermodynamic concepts of entropy and Shannon's Information Theory by methods such as Jayne's Maximum Entropy Principle have been well described \citep{1992Kapur}, and a number of papers support the use of a lognormal function on thermodynamic grounds as an appropriate model to describe systems undergoing information loss \citep{2017MNRAS.467.5022H}.
Theoretical hydrodynamic simulation by \citet{1998PhRvE..58.4501P} suggested that driven turbulence produces a local LN density distribution, and column density observations of star-forming and non-star-forming molecular clouds have supported this \citep{2001ApJ...546..980O, 2009AAp...508L..35K, 2010MNRAS.405L..56B}, with a probability distribution function (PDF) that resembles a lognormal function and a high mass tail attributed to turbulence and self-gravity \citep{2014AAp...571A..95F}.
\citet{2010AAp...511A..85P} measured the entropy profiles of 31 nearby galaxy clusters selected on X-ray luminosity, without morphological bias. 
The observed distributions showed a centrally concentrated excess entropy extending to larger radii in lower mass systems, with a large dispersion in scaled entropy in the inner regions, possibly accounted for by cool cores and dynamical activity but becoming increasingly self-similar at large radii. 
\citet{1994PhDT..Pichon} used perturbation theory to analyse bi-symmetric instability generated by a structure such as a bar. 
He derived a distribution function corresponding to the extremum of entropy, given some supplementary constraints such as linearity in the perturbation, and concluded that a state of maximum entropy compatible with total energy and angular momentum conservation corresponds to uniform rotation. 
\citet{2017MNRAS.467.5022H} used a radial migration model to generate a state close to maximum-entropy. By assuming circularity of the orbits and a maximum entropy distribution of angular momentum, they showed that the derived surface density varied as $\exp(R/R_0)^{1/2}$ at large radii and as $R^{-1}$ at small radii for a model with no halo.

In this paper we use a lognormal function with Newtonian gravity to describe the disc density distribution.
This can mimic a wide range of RCs with good accuracy \citep{2015MNRAS.448.3229M} using the general mathematical form modified to the more physical form of Eqn.~\ref{eq:LN2}:
\begin{equation}
\Sigma(r)=\frac{\Sigma_0}{r/r_\mu}\exp{\left(-\frac{[\log(r/r_\mu)]^2}{2\sigma^2}\right)}\,,
\label{eq:LN2}
\end{equation}
where $\Sigma(r)$ is the disc surface density ($M_\odot$~kpc$^{-2}$); $r$ is the normalized radial variable (kpc); $r_\mu$ is the radial scale factor (kpc); $\sigma$ is the standard deviation of the natural logarithm of the radius; and $\Sigma_0$ is the surface density of the disc at $r=r_\mu$ ($M_\odot$~kpc$^{-2}$).
\begin{figure}
   \centering
	\includegraphics[width=0.75\textwidth]{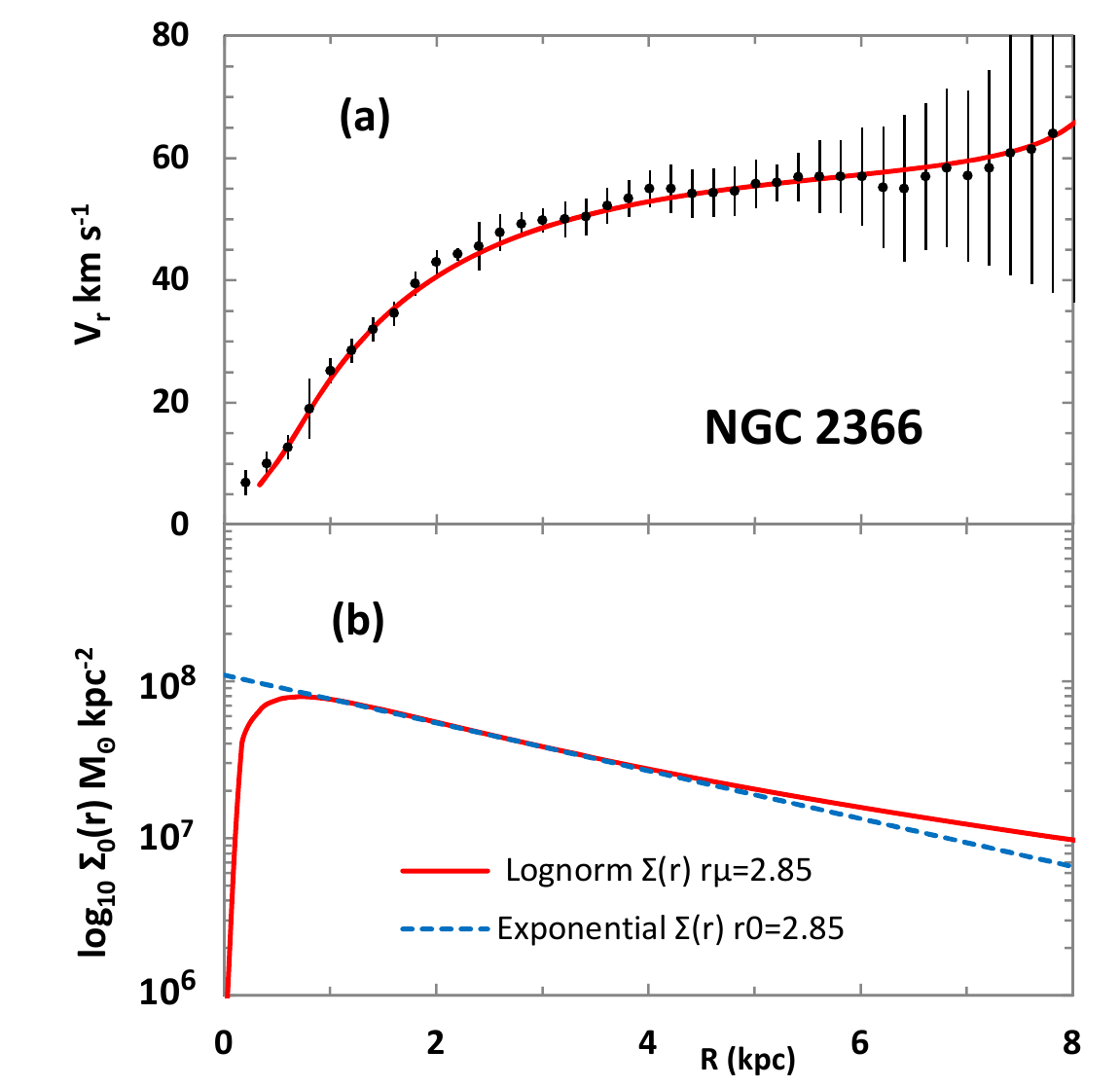}
   \caption{NGC~2366 (a) rotational velocity for the LN model (solid line) and observations \citep{2008AJ....136.2648D}. (b) Radial distribution of density ($M_\odot$/kpc$^2$) for the LN model (solid line) and Freeman exponential disc model (dashed line)} 
   \label{fig:NGC2366}
	\vspace*{8pt}
\end{figure}

The lognormal model has three independent parameters. $\Sigma_0$ only changes the scale of the rotational velocity, $V(r)$, by sliding the curve up or down with no change in shape, but varying $\sigma$ and $r_\mu$ provides a variety of possible curves that demonstrate typical D--, F-- and R--type curves that are generally consistent with those described by \citet{2001ApJ...563..694V}.
The assumed disc radius, $R_{max}$, is not a free parameter, as it can always be normalized to unity. Measuring the actual disc radius, however, is non-trivial, as it is often unclear where the disc terminates, or if it  truncates abruptly. 
Conventionally, the half-light radius may be used, but for the LN model RCs, $R_{max}$ was generally defined as the last reported observation, plus one half-bin size, using the quoted values in kpc or converting from arc-secs using the reported distance parameter \citep{2015MNRAS.453.2214M}.

The LN distribution fulfils the probability distribution for disc systems in a number of important ways: the radius where the stars orbit must be $>0$, the distribution is highly skewed, and normalization of the function to unity corresponds to the probability that the disc contains all the angular momentum of the galaxy.
It is also smoothly asymptotic to zero at the core rather than peaking to a cusp where rotation is unsupported (Eq.~\ref{eq:LN2}).
The LN function does not describe the bulge which is assumed to be non-rotating, and the gravitational potential of which is additive to the total potential within the disc.
The high angular momentum of the discs and their internal density and velocity profiles resemble hurricanes, and this similarity extends to the null central velocity required for spin as emphasised by Criss and Hofmeister \citep{2018Galax...6..115C}.

\begin{figure}
   \centering
	\includegraphics[width=0.6\textwidth]{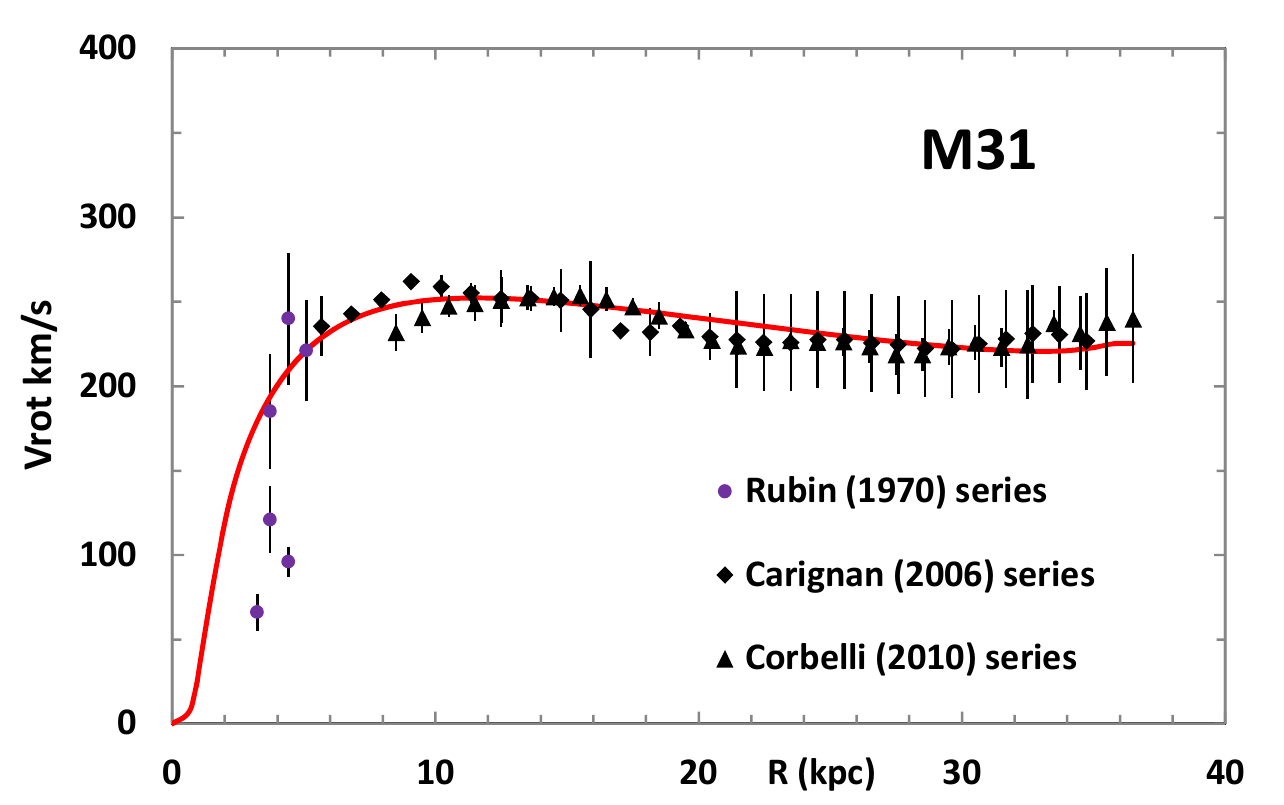}
   \caption{RC for the massive galaxy M31, with rotational velocity for the LN-model (solid line) and observations with error bars (\citep{1970ApJ...159..379R, 2006ApJ...641L.109C, 2010AA...511A..89C}). } 
   \label{fig:M31}
	\vspace*{8pt}
\end{figure}

In contrast to an exponential distribution, these characteristics match the rotation curves for both dwarf galaxies such as NGC~2366 and large galaxies such as M31 (Figs.~\ref{fig:NGC2366}(a) and \ref{fig:M31} respectively), and satisfy the observed disc density distributions for many spiral galaxies \citep{2012msma.book.....F}.
Stopping the surface density abruptly at $R_{max}$ produces a noticeable terminal rise in the RC \citep{2015MNRAS.448.3229M}, but although Eqn.~\ref{eq:LN2} is exact only in the limit $r\rightarrow\infty$, in practice $\Sigma(r)\rightarrow0$ as $r\rightarrow R_{max}$, the maximum radius for observations (kpc) beyond which gas and dust at the galactic periphery become undetectable. 
Although the terminal density may fall away more gradually, it is difficult to detect this termination observationally because any observable matter will already be included in the disc; matter beyond the detectable disc boundary will by definition be unobserved. 
Nevertheless, some observers have reported $\hh$ observations that showed no evidence of stopping at their limit of detection, and -- with increasing sensitivity of observations -- there is now evidence for some $\hh$ and molecular gas components extending beyond the original disc boundaries, usually described by adding further exponential components to the disc boundary as a biaxial or triaxial component to the disc \citep{2008AJ....135...20E, 2013AJ....146..104H}. 

The ability of the LN model to generate realistic RCs with a good fit to observations is demonstrated for the dwarf galaxy NGC~2366 in Fig.~\ref{fig:NGC2366}(a), with $\Sigma_0=1.15\times10^8$~M$_\odot$/kpc$^2$, $\sigma=1.16$, $r_\mu=2.85$~kpc, $R_{max}=8.3$~kpc.
The high-resolution data is taken from the THINGS survey \citep{2008AJ....136.2761O}. 
This survey enabled the effects of random non-circular motions due to collapsing gas clouds in star formation processes, bars, spiral density waves, and warps in the disc to be minimized with the construction of a ``bulk'' velocity field showing the underlying undisturbed rotation, as described by \citet{2008AJ....136.2761O}, with tracers such as $\hh$ and H$\alpha$ defined to follow circular orbits.
NGC~2366 has a relatively flat F-type curve, but shows a strong terminal rise beyond 7.5~kpc. 
This is a feature of many RCs with an abrupt termination \citep{2008AJ....135...20E}, and the fitted LN curve fits the observations well, including the terminal rise.

Fig.~\ref{fig:NGC2366}(b) also compares the NGC~2366 LN density curve used to generate the RC with the best-fitting exponential density curve. 
The LN curve (red solid line) is seen to overlie the exponential curve (blue dashed line) for some portion of the radial distance, implying a possible mechanism for the S\'{e}rsic exponential models. 
Observations show there are more stars than expected in a Freeman disc at small radii where bulge stars predominate near the galactic centre, whereas in the LN model $\Sigma(r)\rightarrow0$ as $r\rightarrow0$ reflecting the collapse of the rotation curve where rotation is unsupported and the disc disappears (Fig.~\ref{fig:NGC2366}(b)). 

\section{Generating the spin parameter and virial mass estimator}
\label{sec:VME}
Equation~(\ref{eq:LN2}) is an integrable function allowing the total disc mass to be calculated (Eq.~\ref{eq:LNmass}):
\begin{equation}
M_{disc}=\Sigma_0\sqrt{2\pi^3}r_\mu^2\sigma\exp\left(\frac{\sigma^2}{2}\right) \left[1-\erf\left(\frac{\sigma^2-\log{(R_{max}/r_\mu)}}{\sigma\sqrt{2}}\right)\right]\,,
\label{eq:LNmass}
\end{equation}
and this theoretical disc mass may be compared to observational data for the system. 
Using a standard function to describe the disc mass distribution also enables other parameters -- such as angular momentum and disc total energy -- to be computed for a comparative analysis of other galaxy disc properties, such as the dimensionless spin parameter, $\lambda$ (Eq.~\ref{eq:spin}) \citep{1969ApJ...155..393P, 2000ApJ...538..477N, 2002MNRAS.332..456V, 2015ApJ...812...29T}: 
\begin{equation}
\lambda\equiv\frac{J|E|^{1/2}}{GM^{5/2}}\,,
\label{eq:spin}
\end{equation}
where $M$ is the total gravitational disc mass, $J$ is total angular momentum, and $|E|$ is total energy of the disc, computed using the LN model for a sample of 38 galaxies of varying morphologies using observational RC data, with radii ranging from 3.0--130 kpc and disc masses spanning more than three decades (Fig.~\ref{fig:AngMom}).
These gave a mean value of $\lambda\simeq 0.423\pm0.014$ \citep{2015MNRAS.453.2214M}, comparable to Binney \& Tremaine's theoretical value for the exponential disc, $\lambda=0.425$ \citep{2008Binney}, implying that $\lambda$ is a universal function and the LN function is a valid description of the disc as a system of maximum entropy. 

\subsection{Changes in Mass and Angular Momentum to Reach Equilibrium}
\label{sec:AngMom}
\begin{figure}
   \centering
	\includegraphics[width=0.75\textwidth]{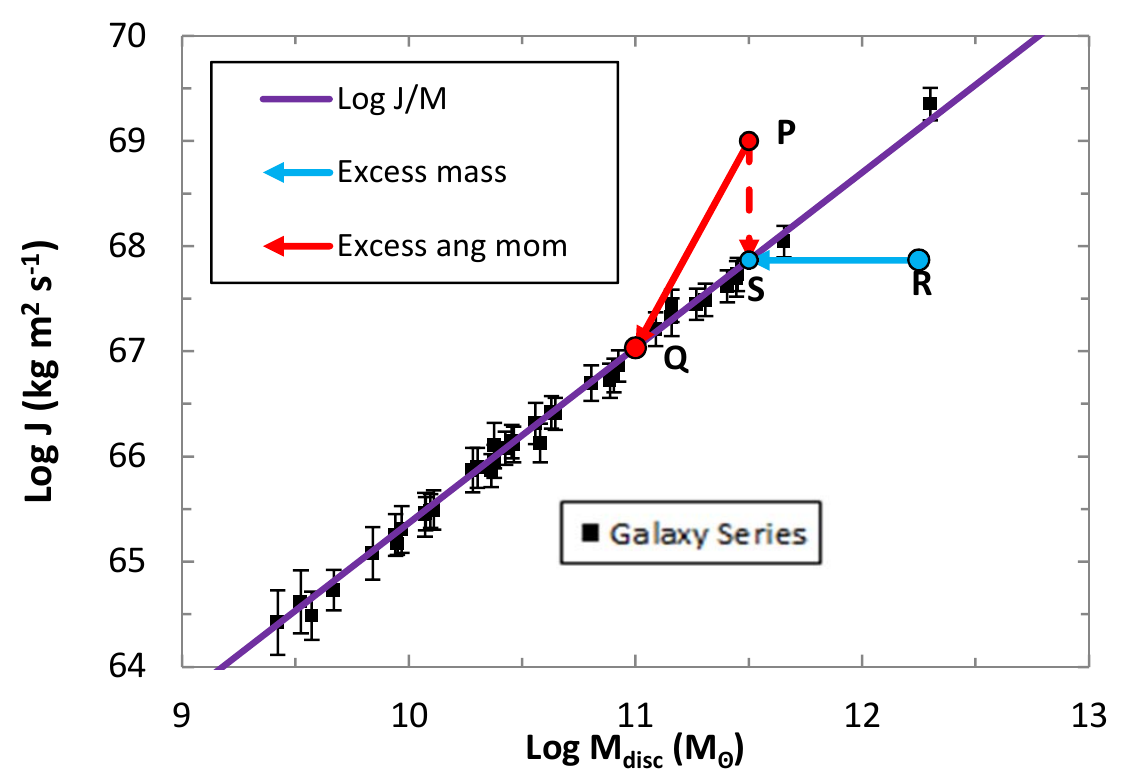}
   \caption{Log-log plot of angular momentum {\em v.} disc mass for the LN-model galaxies (squares) and the theoretical slope 5/3 (purple solid line). Idealised evolutionary changes are shown for a disc with excess mass dumping mass into the core while conserving angular momentum (R--S); and a disc with excess angular momentum shedding mass and angular momentum to a satellite galaxy, leaving a less massive disc (P--Q).} 
   \label{fig:AngMom}
	\vspace*{8pt}
\end{figure}
A log plot of $J$ {\em v.} $M$ for the LN model, using the derived disc masses, is shown in Fig.~\ref{fig:AngMom}, with r.m.s. best-fitting slope of $1.683\pm0.018$, which correlates well with the theoretical slope of $5/3$  \citep{1980MNRAS.193..269V}.
It is, however, unlikely that the initial total induced angular momentum/unit mass of the proto-disc would have had this exact spin parameter, and two scenarios are shown: 
(a) excess $M$ over $J$ (point R); 
and (b) excess $J$ over $M$ (point P). 
To reach stability, each of these must move to the ($J/M$) line as suggested in Fig.~\ref{fig:AngMom}. 
In case (a), it is possible that particles with low angular momentum will fall towards the centre of the proto-galaxy by a process of core-dumping; in effect, the proto-disc will lose mass to the bulge (line R--S).
In case (b), it is unlikely that the proto-disc can lose pure angular momentum (line P--S), but it may spin-off mass and angular momentum together to move to point $Q$ on the ($J/M$) line, as shown (line P--Q). 

Many disc galaxies possess a number of satellite galaxies, and this spin-off mass may go on to form a number of distinct and separate orbiting satellite galaxies, removing both mass and angular momentum from the primary disc.
A possible observational consequence of this relationship is that the number of satellite galaxies possessed by a disc system might be inversely proportional to the mass of its bulge.
Additional evidence for migration is the exceptionally low metallicities of some nearby molecular clouds compared to the average metallicity of the local interstellar medium and the high metallicity of the Sun \citep{2002MNRAS.336..785S}, suggesting that these clouds formed when the proto-galaxy was forming.

\section{The Mass Discrepancy Relation (MDR)}
\label{comparisons}
Calculation of  the theoretical total dynamic masses ($M_{dyn}$) computed with a LN density distribution allows direct comparison with the observational baryonic mass (stars+gas+other components, $M_{bar}$) (Table~\ref{table:masses}). 
Total observational disc masses were estimated by summing the components of luminosity mass and computed hydrogen and helium gas mass from published values for $\hh$, although accurate determination of the total baryonic mass distribution remains difficult \citep{2008AJ....136.2563W} and even for the Milky Way there are large intrinsic uncertainties \citep{2010ApJ...720L.108G, 2014ApJ...785...63B}.
As with all values for $M_{gas}$ in Table~\ref{table:masses}, the values for $\hh$ mass were multiplied by a factor 1.4 to account for the presence of helium, but dust and molecular and ionized gas are not quantified in the mass models, although for neutral gas this can be derived directly by integrating the $\hh$ map.

The problem of calculating $M^*$ is non-trivial, requiring knowledge of the Initial Mass Function (IMF) and the stellar mass-to-light ratio ($\Upsilon^*$). 
This itself depends on several poorly constrained factors including age, colour, metallicity, dust extinction, and recent star formation, and Tutukov has emphasized how this will change with galactic evolution because $\Upsilon^*$ was almost certainly lower in the past when the star formation rate (SFR) was higher and there was less obscuration by dust \citep{2000ARep...44..711T, 2016NatAs...1E...3P}.
Unfortunately these factors are interdependent, giving $\Upsilon^*$ a large uncertainty in the mass models, leading many studies to assume a min-max disc approach \citep{1986Albada, 2008AJ....136.2648D}, with a minimum disc mass from assuming a majority of the rotation arises from the DM halo, and the maximum disc hypothesis providing an upper limit on  $\Upsilon^*$ by maximizing the rotation contribution of the stellar disc.
Mass observational errors in Table~\ref{table:masses} were taken from quoted values where available, or estimated from uncertainties in the $\hh$ maps and the mass-light ratios such as those cited by \citet{2012AJ....143...40M}. 
\begin{figure}
   \centering
	\includegraphics[width=0.6\textwidth]{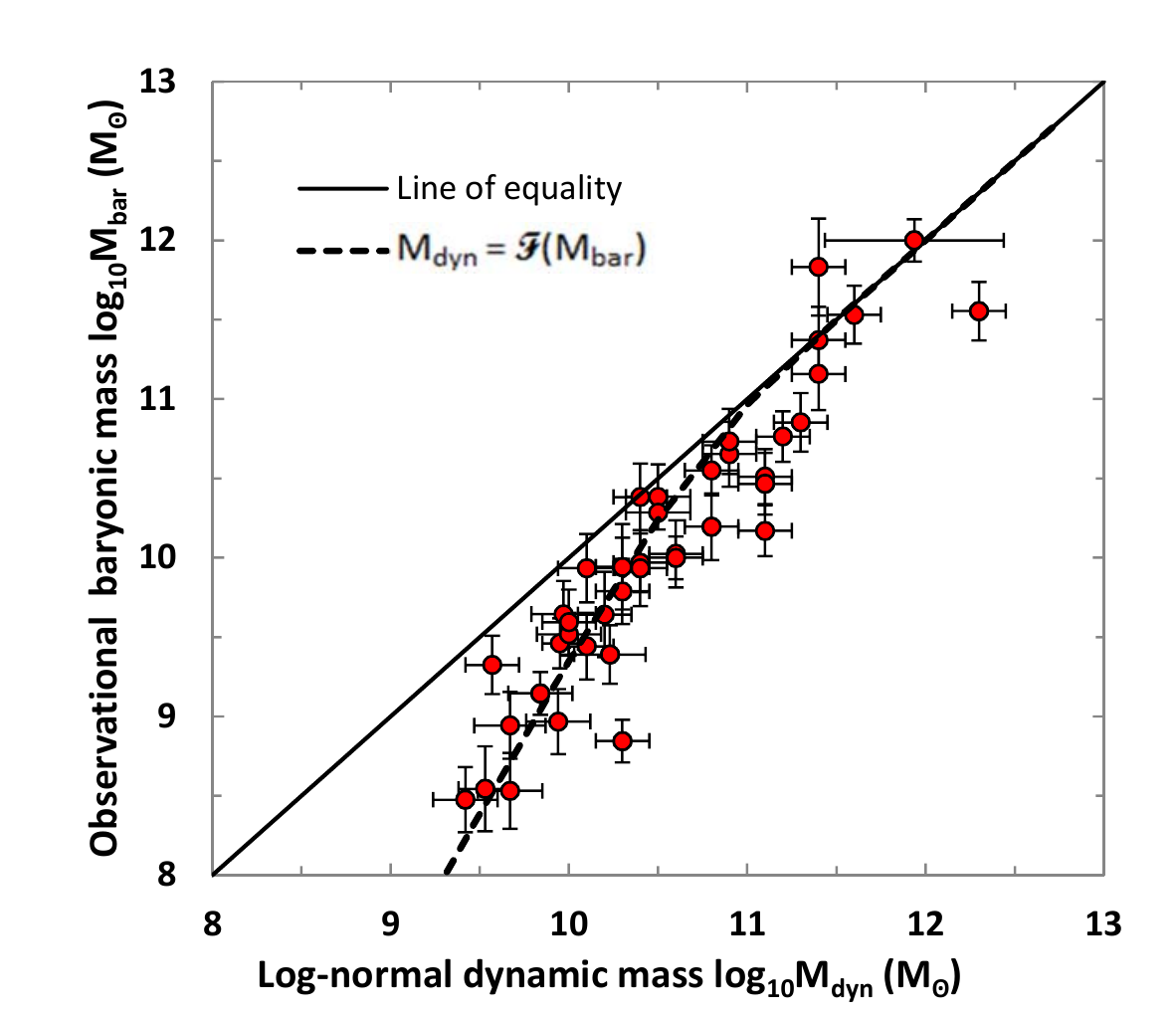}
   \caption{Log-log plot of the modelled lognormal dynamic disc mass {\em vs.} the estimated observational baryonic masses ($M_g+M^*$) for a wide range of galaxy masses and types. The solid line is the line of mass equality; the dashed line is for $M_{dynamic}=f(M_{baryonic})$, with $M_{\dagger}=3.98\times10^{10}~M_{\odot}$ (see text).} 
   \label{fig:Mass_comparisons}
	\vspace*{8pt}
\end{figure}

The total observational baryonic disc masses ($M_{bar}$) calculated from $M_{gas}+M^*$ for 41 galaxies widely spaced in mass and type are plotted against their corresponding LN dynamic masses ($M_{dyn}$) for comparison in Fig.~\ref{fig:Mass_comparisons}.
The disc masses generated by the LN distribution demonstrate an inverse mass deficit relation (MDR), with an increasing mass deficiency with decreasing disc mass. 
This confirms a discrepancy that has been described as the mass discrepancy acceleration relation (MDAR) by \citet{2016MNRAS.461.2367J}, and convincingly demonstrated by \citet*{2016PhRvL.117t1101M} who plotted 2693 points in terms of the local acceleration across the discs of 153 galaxies to show a strong Radial Acceleration Relation (RAR) with a one-fit parameter, the acceleration scale, $g_\dagger=1.20\pm0.26\times10^{-10}$~m~s$^{-2}$  \citep{2016PhRvL.117t1101M}, where the mass discrepancy becomes pronounced.
The apparent increase in mass discrepancy with decreasing disc mass in Fig.~\ref{fig:Mass_comparisons} may be correlated with the RAR by a corresponding mass scale, $M_\dagger$ (Eq.~\ref{eq:MG}):
\begin{equation}
M_{dynamic}=\frac{M_{bar}}{1-\exp(-\sqrt{M_{bar}/M_\dagger})}\,,
\label{eq:MG}
\end{equation}

This is shown as the dashed line in Fig.~\ref{fig:Mass_comparisons}, with an inflexion point at $M_\dagger=3.98\pm0.54\times10^{10} M_\odot$, that may be compared to McGaugh's acceleration scale inflection point at $g_{\dagger}$. 
Despite wide margins of error in deriving the total observational disc mass, the deficiency in observational baryonic mass to computed theoretical dynamic mass appears to bear a systematic inverse relationship to the computed galactic mass, approaching the theoretical disc mass asymptotically for the more massive systems, confirming the earlier observations \citep{2016PhRvL.117t1101M} and lending further support to the LN model as a satisfactory universal predictor of dynamic mass. 

\begin{table*}
\centering
\caption{Observational Masses and Computed lognormal Masses for 41 Disc Galaxies.}
\label{table:masses}
\begin{tabular}{l c c c c c l}
\hline
Name & D & M$_{gas}$  & M$^{*}$ & Total Mass & LN Mass & Refs \\
 & Mpc & (log$_{10}M_\odot$) & (log$_{10}M_\odot$) & (log$_{10}M_\odot$) & (log$_{10}M_\odot$) & [D][$M_{disc}$] \\
\hline
DDO 154&4.04&8.40&8.00&8.54$\pm$0.27&9.53$\pm$0.15&\citep{2012AJ....143...40M}\citep{1998ApJ...506..125C}\\
F563-V2&61.0&9.51&9.74&9.94$\pm$0.18&10.30$\pm$0.15&\citep{2000ApJ...531L.107S}\citep{2005ApJ...632..859M}\\
F568-1&85.0&9.87&9.50&10.02$\pm$0.21&10.60$\pm$0.15&\citep{2000ApJ...531L.107S}\citep{2005ApJ...632..859M}\\
F568-3&77.0&9.71&9.62&9.97$\pm$0.18&10.40$\pm$0.15&\citep{2000ApJ...531L.107S}\citep{2005ApJ...632..859M}\\
F568-V1&84.8&9.53&9.82&10.00$\pm$0.13&10.60$\pm$0.15&\citep{2012AJ....143...40M}\citep{2005ApJ...632..859M}\\
F574-1&96.0&10.32&9.52&10.38$\pm$0.21&10.40$\pm$0.15&\citep{2000ApJ...531L.107S}\citep{2012MNRAS.421.1273O}\\
IC 2574&3.91&9.20&8.94&9.39$\pm$0.18&10.23$\pm$0.20&\citep{2012AJ....143...40M}\citep{2005ApJ...632..859M}\\
M31&0.78&9.70&11.36&11.37$\pm$0.21&11.40$\pm$0.15&\citep{2006ApJ...641L.109C}\citep{2006ApJ...641L.109C}\\
Malin 1&366.0&10.97&- - -&12.00$\pm$0.13&11.94$\pm$0.50&\citep{2010AAp...516A..11L}\citep{2007AAS...20925202B}\\
Milky Way&- - -&- - -&- - -&11.83$\pm$0.31&11.40$\pm$0.15&[--]\citep{2014ApJ...785...63B}\\
NGC 1705&5.10&8.23&8.23&8.53$\pm$0.24&9.67$\pm$0.18&\citep{2012AJ....143....1E}\citep{2006MNRAS.365..759R}\\
NGC 2366&3.27&8.79&8.41&8.94$\pm$0.21&9.67$\pm$0.20&\citep{2012AJ....143...40M}\citep{2008AJ....136.2648D}\\
NGC 2403&3.16&9.67&10.04&10.20$\pm$0.21&10.80$\pm$0.15&\citep{2012AJ....143...40M}\citep{2005ApJ...632..859M}\\
NGC 2683&8.59&8.70&10.54&10.55$\pm$0.16&10.80$\pm$0.15&\citep{2013AJ....146...86T}\citep{2005ApJ...632..859M}\\
NGC 2841&14.10&10.23&11.51&11.53$\pm$0.18&11.60$\pm$0.15&\citep{2012AJ....143...40M}\citep{2005ApJ...632..859M}\\
NGC 2903&8.90&9.49&10.74&10.76$\pm$0.16&11.20$\pm$0.15&\citep{2012AJ....143...40M}\citep{2005ApJ...632..859M}\\
NGC 2915&3.78&8.78&7.99&8.85$\pm$0.13&10.30$\pm$0.15&\citep{2005ApJ...632..859M}\citep{2005ApJ...632..859M}\\
NGC 2976&3.58&8.53&9.25&9.33$\pm$0.18&9.57$\pm$0.15&\citep{2012AJ....143...40M}\citep{2008AJ....136.2648D}\\
NGC 3198&13.80&9.80&10.36&10.47$\pm$0.19&11.10$\pm$0.15&\citep{2012AJ....143...40M}\citep{2005ApJ...632..859M}\\
NGC 3521&8.00&9.80&10.81&10.85$\pm$0.18&11.30$\pm$0.15&\citep{2012AJ....143...40M}\citep{2005ApJ...632..859M}\\
NGC 3726&13.37&9.79&10.42&10.51$\pm$0.17&11.10$\pm$0.15&\citep{2013AJ....146...86T}\citep{2005ApJ...632..859M}\\
NGC 3741&3.0&8.45&7.24&8.48$\pm$0.21&9.42$\pm$0.18&\citep{2005ApJ...632..859M}\citep{2005ApJ...632..859M}\\
NGC 4217&20.14&9.40&10.63&10.65$\pm$0.21&10.90$\pm$0.15&\citep{2013AJ....146...86T}\citep{2005ApJ...632..859M}\\
NGC 4389&9.42&8.75&9.37&9.46$\pm$0.16&9.95$\pm$0.10&\citep{2013AJ....146...86T}\citep{2005ApJ...632..859M}\\
NGC 6946&5.5&10.43&10.43&10.73$\pm$0.21&10.90$\pm$0.15&\citep{2007AJ....133..791S}\citep{2005ApJ...632..859M}\\
NGC 7331&13.87&10.04&11.12&11.16$\pm$0.23&11.40$\pm$0.15&\citep{2013AJ....146...86T}\citep{2005ApJ...632..859M}\\
NGC 7793&3.38&9.46&9.76&9.93$\pm$0.22&10.10$\pm$0.16&\citep{2013AJ....146...86T}\citep{2005ApJ...632..859M}\\
NGC 925&8.91&10.15&10.01&10.38$\pm$0.21&10.50$\pm$0.18&\citep{2013AJ....146...86T}\citep{2008AJ....136.2648D}\\
UGC 128&58.5&9.96&9.76&10.17$\pm$0.16&11.10$\pm$0.15&\citep{2012AJ....143...40M}\citep{2005ApJ...632..859M}\\
UGC 2885&75.9&10.70&11.49&11.55$\pm$0.18&12.30$\pm$0.15&\citep{2012AJ....143...40M}\citep{2005ApJ...632..859M}\\
UGC 5750&56.1&9.71&9.00&9.79$\pm$0.21&10.30$\pm$0.15&\citep{2012MNRAS.421.1273O}\citep{2012MNRAS.421.1273O}\\
UGC 6399&15.5&8.85&9.32&9.44$\pm$0.21&10.10$\pm$0.15&\citep{1998ApJ...503...97S}\citep{1998ApJ...503...97S}\\
UGC 6446&15.5&9.51&9.07&9.64$\pm$0.27&10.20$\pm$0.15&\citep{1998ApJ...503...97S}\citep{1998ApJ...503...97S}\\
UGC 6667&18.2&8.90&9.40&9.52$\pm$0.13&10.00$\pm$0.18&\citep{2013AJ....146...86T}\citep{2005ApJ...632..859M}\\
UGC 6818&19.5&9.00&8.60&9.15$\pm$0.13&9.84$\pm$0.18&\citep{2013AJ....146...86T}\citep{2005ApJ...632..859M}\\
UGC 6917&15.5&9.53&9.73&9.94$\pm$0.27&10.30$\pm$0.15&\citep{1998ApJ...503...97S}\citep{2005ApJ...632..859M}\\
UGC 6923&18.67&9.07&9.44&9.59$\pm$0.21&10.00$\pm$0.15&\citep{2013AJ....146...86T}\citep{2003AAp...412..633Z}\\
UGC 6969&18.6&8.79&8.49&8.97$\pm$0.21&9.94$\pm$0.18&\citep{2002AAp...388..809B}\citep{2002AAp...388..809B}\\
UGC 6973&36.8&9.38&10.23&10.29$\pm$0.02&10.50$\pm$0.18&\citep{2011MNRAS.413..813C}\citep{1998ApJ...503...97S}\\
UGC 6983&18.6&9.46&9.76&9.93$\pm$0.24&10.40$\pm$0.15&\citep{2009AJ....138..392S}\citep{2005ApJ...632..859M}\\
UGC 7089&15.5&9.40&9.28&9.65$\pm$0.21&9.97$\pm$0.18&\citep{1998ApJ...503...97S}\citep{2003AAp...412..633Z}\\
\hline
\multicolumn{7}{l}{\parbox[t]{11.2cm}{
}}\\
\end{tabular}
\end{table*}

\section{Discussion}
\label{sec:Discussion}
The rotation curves (RC) of galactic discs provide a vital tool for studying the dynamics of distant galaxies and can be measured with considerable accuracy. 
Comparing these with the theoretical RCs generated by the total observable mass of the disc and bulge confirms the discrepancies between them.
The failure of observed baryonic mass to account for these RCs led to the concept of a dark matter (DM) halo whose properties may be adjusted empirically to fit the observations \citep{1991AJ....101.1231C}.
However, the inability of experimentalists to identify any DM candidates has led to the postulate that the Newtonian  gravitational constant varies at weak field strengths to produce the observed RCs as the Modified Newtonian Dynamics (MOND) hypothesis \citep{1983ApJ...270..365M}, with gravitational dynamics becoming non-Newtonian in the limit of low acceleration \citep{1996ApJ...473..117S, 1998ApJ...508..132D, 2002ARAA..40..263S, 2010ApJ...718..380S}.

The ability to generate a universal mass model to describe the observed RCs provides a useful method for analysis of mass distribution in the disc, and a number of methods have been described to generate a universal disc mass density-distribution model that mimics any observed RC \citep{2016JMPh....7..680C, 2018P&SS..152...68H}.
Criss and Hofmeister have used the Virial Theorem to model galactic RCs via their linkage of the rotation rate to the gravitational self-potential ($Ug$) and the moment of inertia of oblate spheroids. 
This allowed galactic mass and volumetric density profiles to be extracted from the velocity and its derivative as functions of equatorial radius, giving a direct, unambiguous, and parameter-free inverse model for rotation curves without DM \citep{2018Galax...6..115C, 2018P&SS..152...68H}. 
A computational method was presented by Feng \citep{2014Galax...2..199F} for determining the mass distribution in a mature spiral galaxy from a given rotation curve.
Their surface mass density profiles predicted an approximately exponential law of decay, quantitatively consistent with the observed surface brightness distributions, and suggested that Newtonian dynamics can adequately describe the observed rotation behaviour of mature spiral galaxies.
similarly, \citet{2014arXiv1406.2401P} show that a broad range of galaxy rotation curves can be explained solely by modeling the distribution of baryonic matter in a galaxy.

The maximum entropy model described in this paper provides a physically plausible rationale for a lognormal (LN) surface density distribution that can account for the observed RCs of a wide variety of disc galaxies varying in type, brightness and mass and generates a reasonable model to establish the dynamic mass of the disc for a wide range of disc masses.
It gives a good match to the observational masses of more massive galaxies while approximating the exponential S\'{e}rsic distribution over much of the disc radius.
The LN model has a universal spin parameter with a highly correlated theoretical mass/angular momentum ratio (Fig.~\ref{fig:AngMom}), suggesting a mechanism by which the disc may stabilize from a proto-disc by dumping excess initial mass to the bulge or shedding excess angular momentum to form a satellite galaxy. 

Several independent sources suggest that neither DM nor MOND are universal requirements.
Stellar kinematics of elliptical galaxies have suggested there are few unambiguous cases where DM is needed to fit the data, and dynamical modeling of the data indicates the presence of little if any dark matter in these galaxies’ halos \citep{2000A&AS..144...53K, 2004IAUS..220...39B}.
Two ultra difuse galaxies, NGC1052-DF2 and NGC1052-DF4, have very low dispersion velocities nearly identical to the expected values from their stellar masses alone \citep{2019ApJ...874L..12D, 2019ApJ...874L...5V}, although the distance to these is is still under debate with missing mass comparable to other LSB galaxies \citep{2019MNRAS.486.1192T}.

The LN model accommodates a scenario in which the missing mass is confined to the disc, and plotting these theoretical dynamic masses of the LN model against the observational masses confirms a mass discrepancy relationship (MDR) that increases with decreasing disc mass, in close agreement with the Radial Acceleration Relation (RAR).
\citet{2016PhRvL.117t1101M} explained the RAR using MOND, but the presence of dark matter (DM) as a halo or disc component cannot yet be excluded \citep{2017MNRAS.471.1841N}, although neither explanation is fully satisfactory on physical grounds and \citet{2016MNRAS.461.2367J}, and \citet{2016MNRAS.456L.127D} have suggested that the increase in MDR associated with low mass discs is neither well described by MOND, nor can it arise from a universal NFW profile as this would require a mass-dependent DM density profile in $\Lambda$CDM \citep{2016MNRAS.456L.127D}.

An alternative possibility may be unobserved baryonic mass in the disc periphery of the faint LSB galaxies.
\citet{Salem_2015} have suggested that the Milky Way (MW) hot halo accounts for $4.3\pm0.9\times10^{10}$~M$\odot$ or roughly 50\% of these baryons, and others have suggested a still larger mass of hot gas \citep{2012ApJ...756L...8G}.
Even for the MW, this hot halo has hitherto been undetectable until recent precise measurements of the movements of satellite galaxies, suggesting that there may be similar undetected hot halos associated with other galaxies, and undetected baryonic mass in the LSB galaxies.
The interstellar medium (ISM) is a mixture of gas and dust remaining from the formation of the galaxy, ejected by stars, and accreted from outside. 
The gas is very diffuse with some in the form of single neutral atoms, some in the form of simple molecules, and some existing as ions.
Its chemical composition is about 91\% hydrogen, 9\% helium.
It is observationally important because spectroscopic emission lines from the gas enable measurements of its mass and dynamics, including rotation curves, making it unlikely to be hidden baryonic mass. 
The total density of dust in the ISM is thought to be considerably less than the gas density, and \citet{2007ApJ...663..866D} suggested that $M_{dust}/M_{(HI+H2)}\approx0.01$. 
The composition of the dust particles is highly variable; grains may vary in size by a factor of $100:1$, but any excess of dust in LSBs is unlikely to account for hidden mass if its presence in LSB galaxies is in a similar ratio to the MW.

The size of a stellar system without a sharp boundary may be characterized by a gravitational radius, $r_g$ \cite{2008Binney}, and for a star of mass $1~M_\odot$, the Oort cloud is thought to extend to approximately $1.5\times10^{13}$~km if taken to approximate the gravitational sphere of influence $r_g$ of a stellar-mass star.
The masses of the Oort clouds surrounding such systems are unknown, therefore we may only estimate possible values from the limited information we have for the Oort cloud of the solar system. 
This may contain $10^{11}-10^{12}$ icy bodies, with a total estimated mass of $10^{25}-10^{26}$~kg and a mean density $\sim2\times10^{-15}$~kg~km$^{-3}$, although in one estimate it may approach 2\% of the solar mass, i.e. $\sim4\times10^{28}$~kg \cite{1986EM&P...36..187M}. 
The density and mass of the background population of exo-Oort cloud objects is also completely unknown \cite{2018ApJ...866..131M}, and again -- with such uncertainty even in the MW -- the proportion and mass of icy bodies within other galactic discs is completely unknown, but they will almost certainly exist. 
It is a reasonable hypothesis that lower mass galaxies with a higher proportion of gas and lower star formation rates may have a correspondingly high ratio of undetectable icy bodies to detectable baryons, and this proportion may increase with decreasing overall mass.

The LN model cannot exclude MOND or a gravitationally bound DM component as causative of the MDR.
However, the good agreement between dynamic and baryonic masses for the RCs of high mass galaxies using a plausible mass-density distribution model suggests that some proportion of the unaccounted-for mass in low mass galaxies may be attributable to uncertainties in mass measurements in the disc peripheries of these low surface brightness (LSB) systems.
This may be explained by the presence of DM in the periphery of the disc itself, but the known difficulty and intrinsic errors in assessing the absolute true mass of these systems suggest that at least some of the deficiency may be baryonic \citep{2016JMPh....7..680C}.
Rather than requiring modification to the law of gravity, or a massive undetectable halo of DM, the possibility that it may be accounted for by undetected baryonic matter remains plausible.


\vspace{6pt} 

\acknowledgments{I thank the anonymous referees for their thoughtful and
constructive reports. 
I am grateful to Stacy McGaugh for discussions on generating rotation curves, Erwin de Blok and the THINGS team who freely provided their data, and Alexander Tutukov for valuable insights into the properties of the galactic disc.}

\conflictofinterests{The author declares no conflict of interest.} 

\bibliographystyle{mdpi}
\bibliography{galaxies3} 

\end{document}